\documentclass[pre,preprint,showpacs]{revtex4}
\usepackage[utf8]{inputenc}
\usepackage{amsmath}
\usepackage{amsfonts}
\usepackage{amssymb}

\begin{document}

\title{Microcanonical equations for the Tsallis entropy}

\author{J. Carrete}
\author{L. M. Varela}
\author{L. J. Gallego}

\affiliation{Grupo de Nanomateriais e Materia Branda,
Departamento de F\'{i}sica da Materia Condensada,
Facultade de F\'{i}sica, Universidade de Santiago de Compostela,
E-15782 Santiago de Compostela, Spain}

\begin{abstract}
  Microcanonical equations for several thermodynamic
  properties of a system, suitable for molecular dynamics
  simulations, are derived from the nonextensive Tsallis entropy
  functional. Two possible definitions of temperature, the usual one
  and a ``physical'' modification which satisfies the zeroth law of
  thermodynamics, are considered, and the results from both choices are
  compared. Results for the ideal gas using the first definition of
  temperature are provided and discussed in relation with the
  canonical results reported in the literature. The second choice leaves most
  formulae unchanged from their extensive (Shannon-Boltzmann-Gibbs)
  form.
\end{abstract}


\maketitle

\section{Introduction}

In the last two decades there has been a great deal of interest in
nonextensive entropies to explain physical phenomena such as
anomalous diffusion, believed to be outside the scope of the
conventional and highly successful Shannon-Boltzmann-Gibbs (SBG)
entropy \cite{Shannon}. Among the various reported nonextensive
functionals, the Tsallis entropy \cite{Tsallis} has received
considerable attention. It is defined as

\begin{subequations}\label{grp:tsallis1}
\begin{equation}
S_q\left\{p_l\right\}=-k_B\sum\limits_l p^q_l\log_q p_l,
\label{eqn:tsallis1}
\end{equation}

\noindent where

\begin{equation}
\log_q x=\lim\limits_{q^\prime\rightarrow q}
\frac{x^{1-q^\prime}-1}{1-q^\prime},
\label{eqn:logq}
\end{equation}
\end{subequations}

\noindent $p_l$ being the probability of each microstate $l$
accessible to the system and $k_B$ the Boltzmann constant.  This
definition can be considered an uniparametric generalization of the
SBG functional, which is recovered when the entropic index $q$ equals
one, as can be easily verified using L'Hôpital rule. A further
generalization, not studied in this paper, replaces $k_B$ with a
generic $k\left(q\right)$ under the weaker constraint
$k\left(1\right)=k_B$.

Most discussions about the Tsallis entropy and the theoretical and
computational results from its aplication have been made in the
context of the canonical ensemble, although a satisfactory definition
of the mean energy has taken some effort to achieve \cite{Tsallis}. As
with ordinary extensive statistics, this ensemble is often more
amenable to theoretical calculations. However, the microcanonical
ensemble is clearly more directly accessible to molecular dynamics
(MD) simulation, which consists simply in the integration of Newton's
Second Law, and as such gives rise automatically to the conservation
of energy, but lacks the intrinsic notion of a thermostat. Thus, it is
necessary to have a formalism which allows for the obtention of the
thermodynamic properties of a system from microcanonical
averages.

Several methods have been devised in order to obtain these formulae
for the SBG functional, adjusted to different needs. For instance, Ray
and Graben \cite{Ray}, guided by didactic concerns, use a method based
on fluctuations which involves some arbitrary definitions and is only
valid for relatively large systems. A more systematic and general
method, based on Laplace transforms, was put forward by Pearson,
Halocioglu and Tiller \cite{Pearson}. However, the use of this
transform is only a method of integration. The present paper applies an
equivalent but more straightforward method to the Tsallis entropy, and
it is structured as follows: in the next section, after reviewing some
well-know features of the microcanonical ensemble, a set of formulae
for thermodynamic properties such as the heat capacities and
compressibility coefficient is developed using the conventional
definition of temperature. These formulae are applied to the
nonextensive ideal gas in order to obtain its thermal and caloric
equations. After discussing an important shortcoming of this
definition of temperature, the previously developed results are
changed in order to reflect a physical definition of
temperature. Finally, the main conclusions are summarized.

\section{Formulae obtained using the ordinary temperature}

The system to be studied consists of $N$ classical particles of mass
$m$, with coordinates $\left(\vec{r}^N,\vec{p}^N\right)$ in
$6N$-dimensional phase space, in a recipient of fixed volume $V$ and
with a total energy $E$, resulting from a Hamiltonian which must be
separable into a kinetic and a potential part which depend only on the
positions and momenta, respectively:

\begin{equation}
E=H\left(\vec{r}^N,\vec{p}^N\right)=E_c\left(\vec{p}^N\right)+U\left(\vec{r}^N\right).
\label{eqn:energy}
\end{equation}

\noindent The probability distribution over the microstates compatible
with these restrictions can be obtained using the maximum entropy
(MaxEnt) method, by maximization of \eqref{eqn:tsallis1} submitted to the
normalization $\sum_l p_l=1$. In this way, it is
trivially obtained that all the microstates compatible with the
specified thermodynamic coordinates are equiprobable. In practice, the
energy of the system can only be determined (and constrained) to be in
an interval $\left[E-\delta E,E\right]$, with $\delta E\ll E$ for a
useful measurement. Thus, the number of microstates available to the
system would be $\Gamma\left(E,V,N\right)-\Gamma\left(E-\delta
  E,V,N\right)$, with:

\begin{equation}
\Gamma\left(E,V,N\right)=\frac{1}{h^{3N}}\int\theta\left[E-H\left(\vec{r}^N,\vec{p}^N\right)\right]d\vec{r}^Nd\vec{p}^N.
\label{eqn:Gamma}
\end{equation}

\noindent The spatial limits of integration are $V^N$. The momenta can
be integrated over the whole $\mathbb{R}^{3N}$. $\theta$ is the Heaviside
step function which restricts the integration to the volume in phase
space where the hamiltonian is less or equal than $E$. Finally, $h^3$ is the
phase volume of an individual microstate, used to adimensionalize the
number of microstates, and must be multiplied by $N!$ in the case of
indistinguishable particles. Both the precise value of $h$ and the
presence of $N$ are, however, irrelevant to the following
discussion. For most systems (specially for moderate and large values
of $N$), $\Gamma$ is a strongly increasing function of $E$, so
$\Gamma\left(E-\delta E,V,N\right)\ll \Gamma\left(E,V,N\right)$ and
the number of microstates available to the system can be approximated
by $\Gamma\left(E,V,N\right)$, and the probability of each of them by
its inverse. Substituting this probability in \eqref{eqn:tsallis1} a
straightforward $q$-generalization of the well-known microcanonical
entropy equation is obtained:

\begin{equation}
S_q\left(E,N,V\right)=k_B\log_q \Gamma\left(E,V,N\right).
\label{eqn:tsallis2}
\end{equation}

In conventional thermodynamics, temperature is defined as the inverse of the
partial derivative of entropy with respect to energy:

\begin{equation}
T=\left(\frac{\partial S_q}{\partial
  E}\right)^{-1}_{V,N}=\frac{\Gamma^q}{k_B\Omega},
\label{eqn:temperature}
\end{equation}

\noindent where

\begin{equation}
\Omega=\left(\frac{\partial \Gamma}{\partial
    E}\right)_{V,N}=\frac{1}{h^{3N}}\int\delta\left[E-H\left(\vec{r}^N,\vec{p}^N\right)\right]d\vec{r}^Nd\vec{p}^N.
\label{eqn:Omega}
\end{equation}

\noindent These equations, relating entropy and temperature with
$\Gamma$ and $\Omega$, are responsible for the differences between
the present development and the conventional extensive one.

The microcanonical average of a general magnitude
$\chi\left(\vec{r}^N\right)$ which depends on the positions of the
particles is expressed in two equivalent ways:

\begin{align}
\left\langle \chi
\right\rangle&=\frac{1}{h^{3N}\Gamma}\int\chi\left(\vec{r}^N\right)\theta\left[E-H\left(\vec{r}^N,\vec{p}^N\right)\right]d\vec{r}^Nd\vec{p}^N
\label{eqn:average1}\\
\left\langle \chi
\right\rangle&=\frac{1}{h^{3N}\Omega}\int\chi\left(\vec{r}^N\right)\delta\left[E-H\left(\vec{r}^N,\vec{p}^N\right)\right]d\vec{r}^Nd\vec{p}^N.
\label{eqn:average2}
\end{align}

\noindent The equivalence is based on the hypothesis that $\Gamma$ is
a strongly increasing function of $E$, so only the contribution to it
from the higher energies, $\Omega\delta E$, must be taken into
account, to a good approximation. In the remaining of this paper, only
the second of these formulae will be used.

In order to obtain a microcanonical formula for any thermodynamic
variable, it is necessary to express it as a function of $\Gamma$ or
its partial derivatives, like it has already been done for
temperature. The pressure, isocoric heat capacity and Grünessein
parameter are likewise straightforwardly written as

\begin{align}
p&=T\left(\frac{\partial S_q}{\partial
    V}\right)_{E,N}=\frac{1}{\Omega}\left(\frac{\partial\Gamma}{\partial V}\right)_{E,N}
\label{eqn:pressure}\\
C_V&=\left(\frac{\partial T}{\partial
    E}\right)^{-1}_{V,N}=\left[T\left(\frac{q\Omega}{\Gamma}-\frac{1}{\Omega}\left(\frac{\partial\Omega}{\partial
        E}\right)_{V,N}\right)\right]^{-1}\label{eqn:cv}\\
\gamma&=V\left(\frac{\partial p}{\partial
    E}\right)_{V,N}=\frac{V}{\Omega}\left[\left(\frac{\partial
      \Omega}{\partial
      V}\right)_{E,N}-p\left(\frac{\partial\Omega}{\partial E}\right)_{V,N}\right]
\label{eqn:grunessein}
\end{align}

\noindent Other variables often used in tables and experiments, but
not so inmediately translated into microcanonical language are the
isobaric heat capacity, coefficient of thermal expansion and
isothermal compressibility:

\begin{equation}
C_p=\left(\frac{\partial E}{\partial T}\right)_{p,N},\alpha=\frac{1}{V}\left(\frac{\partial V}{\partial
    T}\right)_{p,N},\kappa_T=\frac{-1}{V}\left(\frac{\partial V}{\partial
    p}\right)_{T,N}
\end{equation}

\noindent This is because of the fact that the constraints more easily
applied to a system in the laboratory (constant temperature or
pressure) are not those of the microcanonical ensemble (constant
energy or volume). However, the previous variables can be related to

\begin{align}
\alpha_E&=\frac{1}{V}\left(\frac{\partial V}{\partial T}\right)_{E,N}=
\left[VT\left(\frac{q}{\Gamma}\left(\frac{\partial\Gamma}{\partial
        V}\right)_{E,N}-\frac{1}{\Omega}\left(\frac{\partial\Omega}{\partial
      V}\right)_{E,N}\right)\right]^{-1}\label{eqn:alphae}\\
B_E&=-V\left(\frac{\partial p}{\partial
    V}\right)_{E,N}=\frac{V}{\Omega}\left[p\left(\frac{\partial\Omega}{\partial
    V}\right)_{E,N}-\left(\frac{\partial^2\Gamma}{\partial V^2}\right)_{E,N}\right]\label{eqn:be}
\end{align}

\noindent using the following identities, obtained from standard
relations among derivatives:

\begin{align}
\kappa_T&=\frac{1}{B_E+\frac{C_V\gamma}{\alpha_EV}}\label{eqn:derivative1}\\
\alpha&=\kappa_T\left(\frac{\gamma
    C_V}{V}-\alpha_EB_E\right)\label{eqn:derivative2}\\
C_p&=C_V\left(1-\frac{\alpha}{\alpha_E}\right)\label{eqn:derivative3}.
\end{align}

It is necessary to find a way to express the derivatives of $\Gamma$
and $\Omega$ in such a way that they can be calculated directly during
a MD simulation. For this purpose, one can observe that the integration in
\eqref{eqn:Gamma} can be performed the following way due to the form
of the hamiltonian in eq. \eqref{eqn:energy}:

\[
\int\limits_{H\left(\vec{r}^N,\vec{p}^N\right)}d\vec{r}^Nd\vec{p}^N=\int\limits_{U\left(\vec{r}^N\right)\le
    E}\left[\int\limits_{\sum\limits_{i=1}^N\sum\limits_{j=1}^3\frac{p_{ij}^2}{2m}\le
    E-U\left(\vec{r}^N\right)}d\vec{p}^N\right]d\vec{r}^N.
\]

\noindent The inner integral in the right-hand side of this equation
is clearly the volume of a $3N$-dimensional sphere of radius
$R=\sqrt{2m\left[E-U\left(\vec{r}^N\right)\right]}$, which equals
$\frac{\pi^{\frac{3N}{2}}}{\left(\frac{3N}{2}\right)!}R^{3N}$
Substituting this into the previous equation, eq. \eqref{eqn:Gamma} becomes

\begin{equation}
\Gamma\left(E,V,N\right)=\frac{\left(2m\pi\right)^{\frac{3N}{2}}}{\left(\frac{3N}{2}\right)!h^{3N}}\int\limits_{V^N}\theta\left[E-U\left(\vec{r}^N\right)\right]\left[E-U\left(\vec{r}^N\right)\right]^{\frac{3N}{2}}d\vec{r}^N.
\label{eqn:Gamma2}
\end{equation}

\noindent Using an almost identical course of reasoning and the
formula for the surface (instead of the volume) of the $n$-dimensional
sphere, eqs. \eqref{eqn:Omega} and \eqref{eqn:average2} can be
reexpressed as:

\begin{align}
\Omega\left(E,V,N\right)&=\frac{\left(2m\pi\right)^{\frac{3N}{2}}}{h^{3N}\left(\frac{3N}{2}-1\right)!}\int\limits_{V^N}\theta\left[E-U\left(\vec{r}^N\right)\right]\left[E-U\left(\vec{r}^N\right)\right]^{\frac{3N}{2}-1}d\vec{r}^N\label{eqn:Omega2}\\
\left\langle\chi\right\rangle&=\frac{\left(2m\pi\right)^{\frac{3N}{2}}}{\Omega
  h^{3N}\left(\frac{3N}{2}-1\right)!}\int\limits_{V^N}\chi\left(\vec{r}\right)\theta\left[E-U\left(\vec{r}^N\right)\right]\left[E-U\left(\vec{r}^N\right)\right]^{\frac{3N}{2}-1}d\vec{r}^N\label{eqn:average3}.
\end{align}

\noindent These expressions are also valid for the extensive case
(since they do not depend on the choice of entropy functional as long
as the latter preserves the microcanonical distribution) and were already
provided by Pearson et al. \cite{Pearson} using the Laplace transform method.

A first practical result can be extracted from these transformations
by setting $\chi\left(\vec{r}\right)=E-U\left(\vec{r}\right)$ (the kinetic
energy expressed as a function of the positions):

\[
\left\langle E_c\right\rangle=\frac{\left(2m\pi\right)^{\frac{3N}{2}}}{\Omega
  h^{3N}\left(\frac{3N}{2}-1\right)!}\int\limits_{V^N}\theta\left[E-U\left(\vec{r}^N\right)\right]\left[E-U\left(\vec{r}^N\right)\right]^{\frac{3N}{2}}d\vec{r}^N,
\]

\noindent and comparing this to eq. \eqref{eqn:Gamma2}, concluding that
$\left\langle
  E_c\right\rangle=\frac{3N}{2}\frac{\Gamma}{\Omega}$. Combining this
with the definition of temperature given in \eqref{eqn:temperature} an
expression for the average kinetic energy as a function of the
temperature can be formulated:

\begin{equation}
\left\langle E_c\right\rangle=\frac{3N}{2}k_BT\Gamma^{q-1},
\label{eqn:equipartition}
\end{equation}

\noindent which can be considered a $q$-generalization of the theorem
of equipartition of energy. The extensive equation is
straightforwardly recovered in the $q\rightarrow 1$ limit. Equation
\eqref{eqn:equipartition} can be used to calculate the microcanonical
temperature during a simulation, but it suffers from the drawback of
having an explicit dependence on $\Gamma$, an integral not usually
employed directly in the field of MD. $\Gamma$ can be
calculated, for a given set of parameters, from a numerical
approximation of integral \eqref{eqn:Gamma2}. Unfortunately, the
complexity of this calculation grows exponentially with $N$ and the
integrand is highly discontinuous, which may preclude microcanonical
calculations for large systems in the Tsallis formalism. Furthermore,
a magnitude which depends on the absolute entropy seems unphysical.

The equation for the pressure can be obtained by substituting
\eqref{eqn:Gamma2} in \eqref{eqn:pressure} (taking into account for
the derivation that both the limits and the integrand depend on $V$)
and comparing with \eqref{eqn:Omega}, which gives:

\begin{equation}
p=\frac{N\Gamma}{\Omega V}-\left\langle\left(\frac{\partial U}{\partial
    V}\right)_{E,N}\right\rangle=\frac{2}{3}\frac{\left\langle E_c\right\rangle}{V}-\frac{1}{3V}\sum\limits_i\left\langle\vec{\nabla_iU}\cdot\vec{r}_i\right\rangle.
\label{eqn:pressure2}
\end{equation}

By differentiating \eqref{eqn:Omega} with respect to $E$ and
comparing it with the result of substituting
$\chi=E_c^{-1}=\left[E-U\left(\vec{r}^N\right)\right]^{-1}$ into
\eqref{eqn:average3}, one arrives at $\left(\frac{\partial\Omega}{\partial
    E}\right)_{V,N}=\Omega\left(\frac{3N}{2}-1\right)\left\langle
  E_c^{-1}\right\rangle$, and eq. \eqref{eqn:cv} can be rewritten as

\begin{equation}
C_V=\left[T\left(\frac{3Nq}{2\left\langle
        E_c\right\rangle}-\left(\frac{3N}{2}-1\right)\left\langle E_c^{-1}\right\rangle\right)\right]^{-1}.
\label{eqn:cv2}
\end{equation}

To calculate the next variable of interest, $\gamma$, there is only an
ingredient lacking according to eq. \eqref{eqn:grunessein}, the partial
derivative of $\Omega$ with respect to the volume. Using the same
method as for the pressure, this parameter is found to be equal to
$\frac{N\Omega}{V}-\Omega\left(\frac{3N}{2}-1\right)\left\langle
  \left(\frac{\partial U}{\partial
      V}\right)_{E,N}E_c^{-1}\right\rangle$, so

\begin{equation}
\gamma=N-V\left(\frac{3N}{2}-1\right)\left[p\left\langle
  E_c^{-1}\right\rangle+\left\langle\left(\frac{\partial U}{\partial V}\right)_{E,N}E_c^{-1}\right\rangle\right].
\label{eqn:gamma2}
\end{equation}

No new derivatives are needed to rewrite expression \eqref{eqn:alphae}
for $\alpha_E$:

\begin{equation}
\alpha_E=\left[VT\left(\frac{3Nqp}{2\left\langle
        E_c\right\rangle}+\left(\frac{3N}{2}-1\right)\left\langle
\left(\frac{\partial U}{\partial V}\right)_{E,N}E_c^{-1}\right\rangle-\frac{N}{V}\right)\right]^{-1}.
\label{eqn:alphae2}
\end{equation}

Finally, differentiating \eqref{eqn:Gamma2} two times with respect to
volume

\[
\frac{1}{\Omega}\left(\frac{\partial^2\Gamma}{\partial
    V^2}\right)_{E,N}=\frac{Np}{V}-\frac{2}{3}\frac{\left\langle
    E_c\right\rangle}{V^2}+\frac{N}{V}\left\langle\left(\frac{\partial
      U}{\partial
      V}\right)_{E,N}\right\rangle+\left\langle\left(\frac{\partial^2U}{\partial
      V^2}\right)_{E,N}\right\rangle-\left(\frac{3N}{2}-1\right)\left\langle\left(\frac{\partial
      U}{\partial V}\right)^2_{E,N}E_c^{-1}\right\rangle
\]

\noindent which, after substitution in \eqref{eqn:be}, gives the
formula for $B_E$:

\begin{align}
B_E=&V\left[\left(\frac{3N}{2}-1\right)\left(
\left\langle\left(\frac{\partial^2 U}{\partial
    V^2}\right)_{E,N}E_c^{-1}\right\rangle
-p\left\langle\left(\frac{\partial U}{\partial
    V}\right)_{E,N}E_c^{-1}\right\rangle\right)\right.\nonumber\\
&\left.+\frac{2}{3}\frac{\left\langle
      E_c\right\rangle}{V^2}-\frac{N}{V}\left\langle\left(\frac{\partial U}{\partial
    V}\right)_{E,N}\right\rangle-\left\langle\left(\frac{\partial^2 U}{\partial
    V^2}\right)_{E,N}\right\rangle\right].
\label{eqn:be2}
\end{align}

\noindent Partial derivatives with respect to volume can easily be
transcribed using gradients as it was done in eq. \eqref{eqn:pressure2}.

These particular expressions have been chosen so all of them except
\eqref{eqn:equipartition} are formally independent of $\Gamma$; hence the
only additional difficulty for a nonextensive simulation is the
calculation of the temperature, which has already been discussed. All
the formulae either are independent of $q$, and thus valid also for
the extensive case, or recover the SBG expression in the $q\rightarrow
1$ limit.

The ideal gas and other simple systems are often used as examples in
textbooks because their phase-space volumes can be calculated
analytically. In particular, for the perfect gas, it is possible to
take $U=0$, so $E\ge 0$ and the integrals over real space in the
previous section become trivial:

\begin{subequations}\label{grp:ideal}
\begin{align}
\Gamma&=\frac{\left(2m\pi
    E\right)^{\frac{3N}{2}}V^N}{h^{3N}\left(\frac{3N}{2}\right)!}\label{eqn:ideal:1}\\
\Omega&=\frac{\left(2m\pi
    \right)^{\frac{3N}{2}}E^{\frac{3N}{2}-1}V^N}{h^{3N}\left(\frac{3N}{2}-1\right)!}\label{eqn:ideal:2},
\end{align}
\end{subequations}

\noindent results which can be substituted in eq. \eqref{eqn:temperature}
to obtain the caloric equation of state:

\begin{equation}
E=\left\{
\left(\frac{3N}{2}\right)^qk_BT\left[\frac{
\left(\frac{3N}{2}-1\right)!
}{
\left(2m\pi\right)^{\frac{3N}{2}}V^N
}
\right]^{q-1}
\right\}^{\frac{1}{\frac{3N}{2}\left(q-1\right)+1}}.
\label{eqn:caloric}
\end{equation}

\noindent Using \eqref{eqn:pressure2} and \eqref{eqn:equipartition}, the
thermal equation of state can likewise be obtained:

\begin{equation}
p=\frac{Nk_BT}{V}\left[
\frac{\left(2m\pi\right)^{\frac{3N}{2}}V^N}{h^{3N}\left(\frac{3N}{2}\right)!}
\right]^{q-1}
\left\{
\left(\frac{3N}{2}\right)^qk_BT\left[
\frac{\left(\frac{3N}{2}-1\right)!}
{\left(2m\pi\right)^{\frac{3N}{2}}V^N}
\right]^{q-1}
\right\}^{\frac{q-1}{q-1+\frac{2}{3N}}}.
\label{eqn:thermal}
\end{equation}

\noindent It is easy to see that as $q\rightarrow 1$ these equations
tend to the well-known extensive ones. Important simplifications arise also
when $N$ is large. These two limits, however, do not
commute. More about this can be found in the canonical treatment by
Abe in \cite{Abe0}. A different microcanonical treatment has been
proposed by Parvan \cite{Parvan}, who includes an extensive thermodynamical
variable $\zeta=\frac{1}{q}-1$ to the description of a system in order
to recover the extensivity in the thermodynamic limit.

\section{Formulae obtained using the physical temperature}

The consideration of the difficulty of obtaining $\Gamma$ by
performing the integration in eq. \eqref{eqn:Gamma2} may suggest the idea
of setting the system at a given temperature by putting it in contact
with a thermostat before starting the microcanonical simulation, so
both $T$ and $\left\langle E_c\right\rangle$ are known and $\Gamma$
can be calculated using eq. \eqref{eqn:equipartition}. However, this
approach fails because the condition of thermal equilibrium between
two systems does not imply that they are at the same temperature
defined by eq. \eqref{eqn:temperature}, as discussed by Abe et al. in
Ref. \cite{Abe}. Thus this definition can be rejected as
unphysical. The correct physical temperature can easily be shown to
be

\begin{equation}
T_{phys}=\left(1+\frac{1-q}{k_B}S_q\right)T=\Gamma^{1-q}T,
\label{eqn:tphys}
\end{equation}

\noindent which has as an inmediate consequence, taking equations
\eqref{eqn:temperature} and \eqref{eqn:equipartition} into account,
that this physical temperature, as a function of phase space
integrals, is independent of $q$, as noted by Toral
\cite{Toral}. Thus, this temperature can be seen as the usual
derivative of the Boltzmann entropy. This has led Gross \cite{Gross}
to conclude that the Boltzmann entropy is enough to describe every
hamiltonian system at equilibrium, extensive or not. This
conclusion has been disputed by Wang \cite{Wang}, and is not applicable
to the aforementioned Parvan formalism \cite{Parvan}. However, a
detailed discussion is clearly outside the objectives of this paper.

With this definition, the equipartition theorem recovers its familiar form:

\begin{align}
T_{phys}&=\frac{\Gamma}{k_B\Omega}\label{eqn:tphys2}\\
\left\langle E_c\right\rangle&=\frac{3N}{2}k_BT_{phys}.
\label{eqn:equipartitionphys}
\end{align}

\noindent In particular, eq. \eqref{eqn:tphys2} is identical to the one
obtained using the SBG entropy. Thus, the expressions given by Pearson
for the pressure, isocoric heat capacity and Grünessein parameter, as well
as the auxiliary variables $\alpha_E$ and $B_E$, or equivalently the
$q\rightarrow 1$ ($T\rightarrow T_{phys}$) limits of the ones given
in the previous section, are still valid:

\begin{align}
  p=&-\left(\frac{\partial E}{\partial V}\right)_{E,N}=\frac{Nk_BT_{phys}}{V}-\left\langle\left(\frac{\partial
        U}{\partial V}\right)_{E,N}\right\rangle\label{eqn:pphys}\\
  C_{V,phys}=&\left(\frac{\partial E}{\partial T_{phys}}\right)_{V,N}=k_B\left[\left(\frac{2}{3N}-1\right)\left\langle
      E_c\right\rangle\left\langle
      E_c^{-1}\right\rangle+1\right]^{-1}\label{eqn:cvphys}\\
  \gamma=&V\left(\frac{\partial p}{\partial E}\right)_{V,N}=N-V\left(\frac{3N}{2}-1\right)\left[p\left\langle
      E_c^{-1}\right\rangle+\left\langle\left(\frac{\partial
          U}{\partial
          V}\right)_{E,N}E_c^{-1}\right\rangle\right] \label{eqn:gammaphys}\\
  \alpha_{E,phys}=&\frac{1}{V}\left(\frac{\partial V}{\partial T_{phys}}\right)_{E,N}=\left[VT_{phys}\left(\frac{3Np}{2\left\langle
        E_c\right\rangle}+\left(\frac{3N}{2}-1\right)\left\langle
      \left(\frac{\partial U}{\partial
          V}\right)_{E,N}E_c^{-1}\right\rangle-\frac{N}{V}\right)\right]^{-1}\label{eqn:alphaephys}\\
B_E=&-V\left(\frac{\partial p}{\partial V}\right)_{E,N}=V\left[\left(\frac{3N}{2}-1\right)\left(
\left\langle\left(\frac{\partial^2 U}{\partial
    V^2}\right)_{E,N}E_c^{-1}\right\rangle
-p\left\langle\left(\frac{\partial U}{\partial
    V}\right)_{E,N}E_c^{-1}\right\rangle\right)\right.\nonumber\\
&\left.+\frac{2}{3}\frac{\left\langle
      E_c\right\rangle}{V^2}-\frac{N}{V}\left\langle\left(\frac{\partial U}{\partial
    V}\right)_{E,N}\right\rangle-\left\langle\left(\frac{\partial^2 U}{\partial
    V^2}\right)_{E,N}\right\rangle\right]\label{eqn:bephys}
\end{align}

The only obstacle to simply adopt all the extensive formulae is the
definition of pressure as a derivative of entropy. To keep its
physical meaning, the new definition of temperature requires that
pressure is expressed as

\begin{equation}
p=T\left(\frac{\partial S_q}{\partial V}\right)_{E,N}=\frac{T_{phys}}{1+\frac{1-q}{k_B}S_q}\left(\frac{\partial S_q}{\partial V}\right)_{E,N},
\label{eqn:pphys2}
\end{equation}

\noindent which means that derivatives involving both $T_{phys}$ and
$p$ must be changed in order to reflect this fact. This affects thermodynamic
equations \eqref{eqn:derivative1}, \eqref{eqn:derivative2} and
\eqref{eqn:derivative3}. Since $T_{phys}$ is a function of $S_q$ and
$T$, it is easy to use the chain rule to relate the physical variables
to the unphysical ones for which equations have already been developed
in the previous section:

\begin{align}
\alpha_{phys}&=\frac{1}{V}\left(\frac{\partial V}{\partial
    T_{phys}}\right)_{p,N}=\left[
\frac{1}{\alpha}+\frac{1-q}{k_B}\left(p+\frac{B_E}{\gamma}\right)V\right]^{-1}\label{eqn:alphaphys}\\
\kappa_{T,phys}&=\frac{-1}{V}\left(\frac{\partial V}{\partial
    p}\right)_{T_{phys},N}=\alpha_{phys}\left[\frac{\kappa_T}{\alpha}+\frac{1-q}{k_B}\left(1+\frac{1}{\gamma}\right)V\right]\label{eqn:kappatphys}\\
C_{p,phys}&=\left(\frac{\partial E}{\partial T_{phys}}\right)_{p,N}=
\left[\frac{1}{C_p}+\frac{1-q}{k_B}\left(1+\frac{p\gamma}{B_E}\right)\right]^{-1}\label{eqn:cpphys}.
\end{align}

\noindent These magnitudes are not independent of $q$ and ultimately
depend on $T$ through $C_V$, which means that knowledge of $\Gamma$ is
still necessary to completely determine the thermodynamics of the system.

\section{Conclusions.}

The thermodynamic behaviour of a system described by the Tsallis entropy
has been studied in the microcanonical ensembles. Two possible
definitions of temperature have been considered, resulting in two
different sets of formulae. These microcanonical formulae can be used
to obtain thermodynamic information (heat capacities, pressure,
dilatation coefficients, etc.) from mechanical averages available
during a MD simulation. However, if the first
definition of temperature is used, which amounts to a direct
generalization of the conventional one, these thermodynamic magnitudes
depend on phase-space integrals which are not easily calculated. The
second definition of temperature is designed to satisfy the zeroth law
of thermodynamics and addresses this problem but only partially,
returning some formulae to their extensive ($q=1$) form. This is
explained by the fact that the physical temperature is, for any value
of $q$, a derivative of the SBG entropy, a relation clearly seen in
the microcanonical formalism.

\textbf{Ackowledgements:} This work was supported by the Spanish
Ministry of Education and Science in conjunction with the European
Regional Development Fund (Grants Nos. FIS2005-04239 and
FIS2007-66823-C02-02) and the Direcci\'{o}n Xeral de I+D+I da Xunta
de Galicia in conjunction with the European Regional Development Fund (Grant
No. INCITE07PXI206076ES). J.C.  wishes to thank the financial support
of the Direcci\'{o}n Xeral de Ordenaci\'{o}n e Calidade do Sistema
Universitario de Galicia, da Conseller\'{i}a de Educaci\'{o}n e
Ordenaci\'{o}n Universitaria-Xunta de Galicia.

\end{document}